\documentclass[aps,preprint,floatfix,nofootinbib,showpacs]{revtex4}
\usepackage{graphicx}
\newcommand{\U}{{\cal U}} 
\begin{document}

\title{Decay of $Z$ Boson into Photon and  Unparticle}
\renewcommand{\thefootnote}{\fnsymbol{footnote}}
\author{ 
Kingman Cheung$^{1,2}$, 
Thomas W. Kephart$^3$,
Wai-Yee Keung$^4$,
and Tzu-Chiang Yuan$^{2}$
 }
\affiliation{$^1$Department of Physics, National Tsing Hua University, 
Hsinchu 300, Taiwan
\\
$^2$Physics Division, National Center for Theoretical Sciences,
Hsinchu 300, Taiwan
\\
$^3$Department of Physics and Astronomy, Vanderbilt University, Nashville,
Tennessee 37235, USA \\
$^4$Department of Physics, University of Illinois, Chicago, Illinois 
60607-7059, USA 
}

\renewcommand{\thefootnote}{\arabic{footnote}}
\date{\today}

\begin{abstract}
We study the decay of the standard model $Z$ boson into unparticle
plus a single photon through a one-loop process.  As in the
anomaly type decay, only the axial-vector part of the $Z$ coupling matching with
the vector unparticle and/or the vector part of the $Z$ coupling
matching with the axial-vector unparticle can give a nonzero
contribution to the decay.  We show that the photon spectrum
terminates at the end point in accord with Yang's theorem.  Existing
data on single photon production at LEP I is used to constrain the
unparticle sector.
\end{abstract}

\pacs{14.80.-j, 12.38.Qk, 12.90.+b, 13.40.Em}

\maketitle

\section{Introduction}

It is well known that the Poincar\'e symmetry group can be enlarged to
the group of conformal symmetry in any number of spacetime dimension
$D$. From the structure of the conformal algebra we learn that
conformal symmetry implies scale invariance, while scale invariance in
general does not necessarily imply conformal symmetry.  It is widely
believed that a local quantum field theory that is scale invariant
will also be conformal invariant for any $D$. Although a formal proof
is still lacking, no counterexample has been found. For $D=2$ it was
first conjectured in \cite{zamolodchikov} and later shown in
\cite{polchinski} that scale invariance does imply conformal symmetry
under broad conditions \footnote{However, a counterexample in the
two-dimensional theory of elasticity was demonstrated in
\cite{cardy}.}.
One of the most important consequences of scale invariance is that the
single particle state must be either massless or have a continuous
mass spectrum. An obvious example for the former case is pure QED,
where the photon is exactly massless.  Minimal coupling of the photon
to charged particles in a gauge invariant fashion guarantees the
photon remains massless to all orders in the perturbation series.
However, particles with continuous mass distributions have been
largely ignored in particle physics due to lack of experimental
evidence.

Recently, an interesting physical possibility for scale invariant stuff 
with continuous mass distribution was
pointed out by Georgi \cite{unparticle}, who coined the term unparticle \footnote{
We use the term ``unparticle" as an uncountable noun, so
just like water, it has no distinct plural.}
to describe a possible scale-invariant hidden sector sitting at an 
infrared fixed point at a high scale $\Lambda_\U$.  
If the hidden sector carries standard model (SM) quantum numbers, it
would be highly constrained by existing experimental data. In Georgi's
scheme \cite{unparticle}, the hidden sector communicates with the SM
content via a messenger sector characterized by a high mass scale $M$.
At energy below $M$, one can integrate out the messenger sector and
end up with the effective operator suppressed by inverse powers of $M$
of the following form
\begin{equation}
\label{effectiveop}
\frac{1}{M^{d_{SM} + d_{UV} - 4}} {\cal O}_{SM} {\cal O}_{UV} \;,
\end{equation}
where ${\cal O}_{SM}$ and ${\cal O}_{UV}$ represent local operators of
the SM and hidden sector with scaling dimensions $d_{SM}$ and $d_{UV}$
respectively.  As one scales down the theory from $M$, the hidden
sector may flow to an infrared fixed point at the scale $\Lambda_\U$
which, for example, can be generated by quantum effects via
dimensional transmutation.  At the fixed point where the hidden sector
becomes scale invariant, the above operator Eq. (\ref{effectiveop})
has to be replaced by a new set of operators of similar form
\begin{equation}
\label{unparticleop}
C_{{\cal O}_{\U}} \frac{\Lambda^{d_{UV} - d_{\U}}_\U}{M^{d_{SM} + d_{UV} - 4}}
 {\cal O}_{SM} {\cal O}_{\U} \;,
\end{equation}
where ${\cal O}_{\U}$ is the unparticle operator with a scaling
dimension $d_\U$ and $C_{{\cal O}_{\U}}$ is the unknown
coefficient. Because the underlying theory is a scale invariant
interacting theory, the scaling dimension $d_\U$ need not have a
canonical value of integer or half-integer, unlike the free boson or
free fermion cases. The unparticle operator ${\cal O}_{\U}$ can be
characterized as scalar, vector, tensor, spinor, etc., according to
its Lorentz group representation.
One prototype \cite{unparticle} hidden sector that can give rise to
unparticle is the weakly interacting Banks-Zaks \cite{banks-zaks}
theory.  Another possibility is the strongly interacting magnetic
phase of certain supersymmetric QCD theories \cite{Seiberg}, as
pointed out in \cite{Fox}. 
A third possibility would
be the hidden valleys model which can be viewed as an unparticle sector 
with a large mass gap \cite{Strassler}.

Even though the scale invariant sector remains unspecified, the
2-point function \cite{unparticle} and the Feynman propagator
\cite{unparticle-propagator,CKY} of the unparticle field operator
${\cal O}_\U$ can be determined by scale invariance.
We note if special {\it conformal
invariance} is imposed it is shown in a recent paper \cite{Grinstein} 
that the form of vector and tensor unparticle propagators 
should be modified. Consequently, 
the polarization sum of the vector and tensor unparticle have to be
modified as well.  However, the new form of the polarization sum does not allow 
one to impose transversality of the vector and tensor unparticle 
unless the scaling dimension $d_\U$ equals to 3.
Many groups have pursued phenomenological studies of 
unparticle physics 
\cite{unparticle,
Fox,
Strassler,
unparticle-propagator,
CKY,
Grinstein,
Luo-Zhu,
Chen-Geng,  
Ding-Yan,
Liao-1,
Aliev-Cornell-Gaur,
Catterall-Sannio,
Li-Wei,
Lu-Wang-Wang,
Greiner,
Davoudiasl,  
Choudhury-Ghosh-Mamta,
Chen-He,        
Mathews-Ravindran,
Zhou,
Liao-Liu,
minimal-walking,
Bander-Feng-Rajaraman-Shirman,
Rizzo,
ungravity,
Chen-He-Tsai,
Zwicky,
Kikuchi-Okada,
Mohanta-Giri,
Huang-Wu,
Lenz,
Choudhury-Ghosh,
Zhang-Li-Li,
LLWT,
Nakayama,
Desh-He-Jiang,
BGHNS, 
Delgado-Espinosa-Quiros,
colored-unparticle,
Neubert,
Hannestad-Raffelt-Wong,
HMS,
Desh-Hsu-Jiang,
Kumar-Das,
BCG,
Liao-2,
Majumdar,
Alan-Pak-Senol,
Freitas-Wyler,
GOS,
Hur-Ko-Wu,
Anchordoqui-Goldberg,
Majhi, 
McDonald,
KMRT,
DMR,
Kobakhidze,
Balantekin-Ozansoy,
Aliev-Savci,
LKQWL,  
Iltan, 
CHLTW,
Lewis, 
AKTVV,
CHHL, 
Sahin-Sahin,
Alan,
CLY,
Huitu-Rai,
Dutta-Goyal,
Mureika,
Cakir-Ozansoy,
Wu-Zhang,
Kikuchi-Okada-Takeuchi,
He-Pakvasa,
Chen-Kim-Yoon,
Bashiry,
Feng-Rajaraman-Tu},
while more conceptual aspects of unparticle were
explored by others 
\cite{Stephanov,Krasnikov,HEIDI,Ryttov-Sannino,JPLee,Holdom,Licht}.
Unparticle does not have a fixed invariant mass, but instead has a
continuous mass spectrum.
Thus, like a massless particle, the unparticle has no rest frame.
This implies that real unparticle is stable and cannot decay.
Direct signals of unparticle can nevertheless be detected
in the missing energy and momentum distributions carried away by the
unparticle once it is produced in a process \cite{unparticle}, while 
virtual unparticle effects can be probed via interference  
with SM amplitudes \cite{unparticle-propagator,CKY}. 

In a previous note \cite{CLY}, Li and two of us 
studied the decay of a SM Higgs ($H$) into vector unparticle  
plus a single photon. We showed that the photon energy 
spectrum for the process is continuously smeared out near its end point
and its branching ratio is comparable to that of the discovery mode 
of $H \to \gamma \gamma$ for an intermediate mass Higgs.
In this note, we study the rare decay of the $Z$ boson into unparticle plus 
a single photon, $Z\to  \U \gamma$, via a triangular loop of SM fermions.  
As in the Higgs decay, the energy spectrum of the photon would have been  
monochromatic had the unparticle had a fixed mass.  However,
due to the nature of the continuous mass spectrum of the unparticle, the
resultant photon energy spectrum is also continuous and the shape
depends sensitively on the scaling dimension $d_\U$.  But unlike the case of 
$H \to \U \gamma$, and many other previously
studied cases, the end point of the photon energy spectrum in the decay
$Z\to  \U\gamma$ goes to zero as governed by Yang's theorem.

\section{Decay rate of $Z \to \U \gamma$}

The interaction of spin-1 unparticle $\U$ with a SM fermion $f$ 
can be parameterized by a term in the effective Lagrangian \cite{unparticle,CKY}
\begin{eqnarray}
{\cal L}_{\mathrm{eff}} & \ni & \frac{1}{\Lambda_\U^{d_\U-1}} \bar f \left( 
\lambda_1^f \gamma_\mu - \lambda^{\prime f}_1 \gamma_\mu \gamma_5 
\right) f O^\mu_{\U} 
\end{eqnarray}
where $ \lambda_1^f $ and $ \lambda_1^{\prime f}$ are the unknown
vector and axial-vector couplings.  
Here we assume the transversality of unparticle operator 
$\partial_\mu O^\mu_\U = 0$ is satisfied and $ O^\mu_\U$ has both vector and axial-vector couplings to the SM fermions.
The process $Z \to \U \gamma$ is induced at one-loop level with 
the standard model fermions circling in a triangle loop diagram. 
The photon always has vector-type interactions with SM fermions.
Possible types of interactions for the $Z-\U-\gamma$ vertex 
are either $AVV$ or $VAV$, where $V (A)$ denotes vector (axial-vector) interaction.
The other two possibilities of $VVV$ and $AAV$ vanish due to Furry's theorem.
Note that the vector couplings of the $Z$ boson are much smaller than the 
axial-vector couplings (at least true for $u$, $d$ and $e$), but we consider 
both types of couplings for the $Z-\U-\gamma$ vertex.

The amplitude square for $Z \to \U \gamma $ can be adapted from an earlier 
calculation in a different context \cite{yee}
\begin{equation}
\sum_{\rm pol} |{\cal M}|^2 = \frac{1}{2 \pi^4} z (1-z)^2 ( 1+z) |{\cal A}|^2 
m_Z^2 \; ,
\end{equation}
where $z = P_\U^2/m_Z^2$. The loop amplitude $\cal A$ is given by
\begin{equation}
\label{loopamp}
{\cal A} =  - \frac{e^2 }{\sin\theta_{\rm w} \cos\theta_{\rm w}}
\, \frac{1}{\Lambda_\U^{d_\U-1}}\, \sum_f \, N_C^f \, Q_f \,
\left( g_A^f \lambda_1^f + g_V^f \lambda_1^{\prime f} \right )\,
{\cal I}\left( z, \eta_f \right ) \;,
\end{equation}
where the color factor $N_C^f=3\,(1)$ for $f$ being a quark (lepton), 
$g_V^f = T_{3f}/2 - Q_f \sin^2\theta_{\rm w}$ and
$g_A^f = T_{3f}/2$ are the vector and 
the axial-vector couplings of the $Z$ boson to the fermion $f$ respectively, 
$Q_f$ is the electric charge of the fermion $f$ and 
$\eta_f = m_f^2 / m_Z^2$.
The loop function ${\cal I}(z,\eta)$ is given by
\begin{equation}
{\cal I}(z,\eta) = \frac{1}{1-z} \left\{
\frac{1}{2} + \frac{\eta}{1-z} \left[ F\left(\eta\right) - F\left(\frac{\eta}{z}\right) \right ]
- \frac{1}{2(1-z)} \left[ G\left(\eta\right) - G\left(\frac{\eta}{z}\right) \right ] \right \} \; ,
\end{equation}
where
\[
F(x) = \left\{ \begin{array}{ll}
   -2 \left( \sin^{-1} \sqrt{\frac{1}{4x}}  \right )^2  & \qquad \qquad 
      \mbox{for $x \ge \frac{1}{4}$}  \\
    \frac{1}{2} \left( \ln \frac{x^+}{x^-} \right )^2 - \frac{\pi^2}{2}
  - i  \pi \ln \frac{x^+}{x^-}  & \qquad \qquad 
      \mbox{for $x < \frac{1}{4}$}      \end{array}
          \right .
\]
and
\[
G(x) = \left\{ \begin{array}{ll}
    2 \sqrt{4x-1} \sin^{-1} \sqrt{\frac{1}{4x}}  & \qquad \qquad 
      \mbox{for $x \ge \frac{1}{4}$} \\
     \sqrt{1-4x}  \left( \ln \frac{x^+}{x^-} - i  \pi \right )
   & \qquad \qquad       \mbox{for $x < \frac{1}{4}$}    
             \end{array}
          \right .
\]
with
\[
  x^\pm = \frac{1}{2} \pm \sqrt{ \frac{1}{4} - x } \;.
\]
The amplitude square vanishes in the limit $z\to 0$ as governed 
by Yang's theorem.

The differential decay width for $Z \to \U \gamma$ is 
\begin{equation}
 d\Gamma = \frac{1}{2 m_Z} \, \overline{\sum} |{\cal M}|^2 \, d\Phi \; ,
\end{equation}
where $\overline{\sum} |{\cal M}|^2 = \frac{1}{3} 
\sum_{\rm pol} |{\cal M}|^2 $ and the phase space factor $d\Phi$ is 
\begin{equation}
 d\Phi = \frac{A_{d_\U}}{16 \pi^2} \,(m_Z^2)^{d_\U-1} \,z^{d_\U -2} \,(1-z)
  \, dz 
\end{equation}
with
\begin{equation}
A_{d_\U} = \frac{16 \pi^{5/2}}{(2 \pi)^{2 d_\U}}
\frac{\Gamma( d_\U + \frac{1}{2} )} {\Gamma(d_\U - 1)\Gamma(2 d_\U)} \; .
\end{equation}
Collecting all the pieces, we have
\begin{eqnarray}
 \frac{d \Gamma}{d z} &=& 
 \frac{e^4 }{192 \pi^6 \sin^2\theta_{\rm w} \cos^2\theta_{\rm w}}\,
\,A_{d_\U} \, m_Z \left( \frac{m_Z^2}{\Lambda_\U^2} \right )^{d_\U-1} \,
 \nonumber \\
&& \times 
\left| \sum_f N_C^f \, Q_f \,
\left( g_A^f \lambda_1^f + g_V^f \lambda_1^{\prime f} \right )\, 
{\cal I}\left(z, \eta_f \right )\right |^2
\, z^{d_\U -1} (1-z)^3 (1+z) \;. \label{dgdz}
\end{eqnarray}
Integrating the above expression over $z$ from 0 to 1, we obtain the
partial width of the channel $Z \to \U \gamma$.

\section{Numerical Results}

\begin{figure}[t!]
\centering
\includegraphics[width=5.8in]{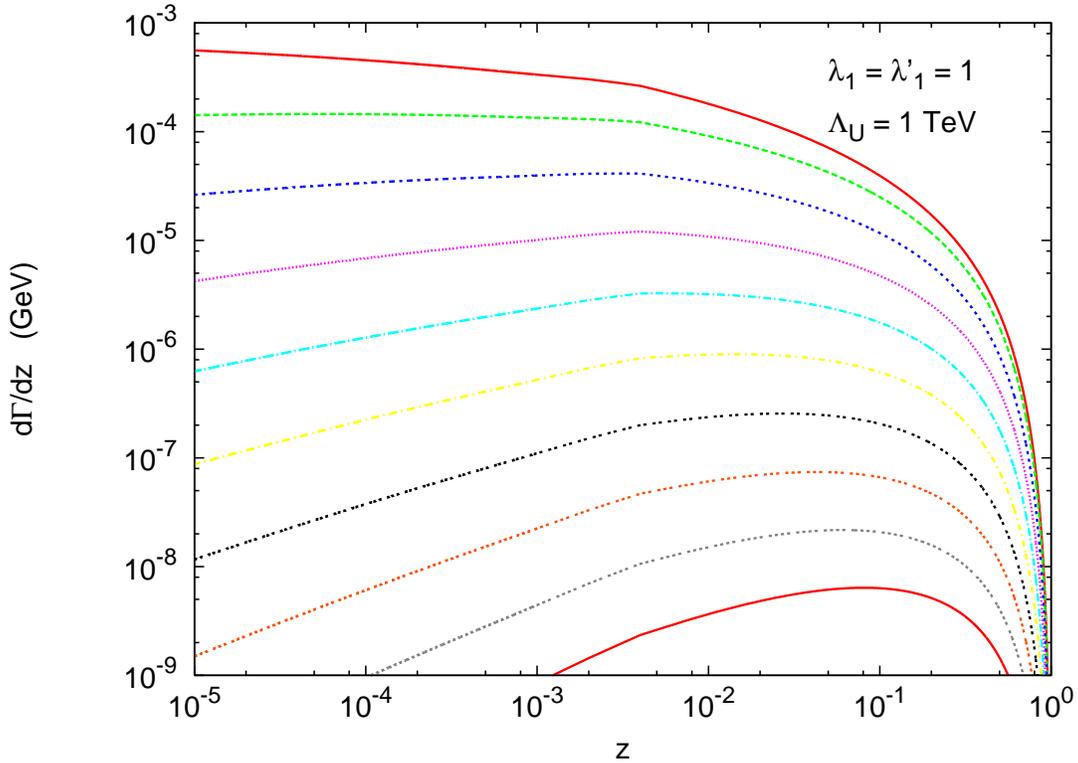}
\caption{\small\label{spectrum}
Spectrum $d\Gamma (Z\to \U \gamma)/dz$, 
where $z = P_\U^2/m_Z^2$, with 
$\lambda^f_1 = \lambda^{\prime f}_1=1$ and $\Lambda_\U=1$ TeV
for $d_\U = 1.1 - 2.0$ (from top to bottom).  
}
\end{figure}
If we ignore the fermion masses as we take $z \to 0$, then the
relevant factor in Eq. (\ref{dgdz}) scales as $z^{d_\U-1} ( 1 + \ln z
)$, where the factor $(1+\ln z)$ comes from ${\cal I}(z,0)$ \cite{yee}:
\begin{eqnarray}
{\cal I}(z, 0) & = & \frac{1}{2 ( 1 - z )} \Biggl( 1 + \frac{\ln \left( z \right)}{1 - z} \Biggr) \; .
\end{eqnarray}  
The photon energy $E_\gamma$ is given by $m_Z
(1 - z)/2$. At $z = 0$, the photon energy reaches its end point and
the unparticle behaves like a massless particle.  Thus, as long as
$d_\U > 1$ the photon energy spectrum vanishes at $z=0$ in accord with
Yang's theorem. 
It is worth mentioned that since $\lambda_1^f$ and $\lambda_1^{\prime f}$ are viewed as 
effective couplings in Georgi's scheme, the following combination
\begin{eqnarray}
&& \sum_f \, N_C^f \, Q_f \, \left( g_A^f \lambda_1^f + g_V^f \lambda_1^{\prime f} \right)
\end{eqnarray}
needs not vanish when summed over SM fermions in contrast with the anomaly-induced decay
of $Z^\prime \to Z \gamma$ in $E_6$ models studied in \cite{yee}.
In Fig.~\ref{spectrum}, we plot the spectrum $d\Gamma/dz$,
where $z = P_\U^2/m_Z^2$, for a range
of $d_\U = 1.1 - 2.0$ (from top to bottom) with 
the democratic assumption of $\lambda^f_1 = \lambda^{\prime f}_1=1$ for all
SM fermions and $\Lambda_\U$ is set to be 1 TeV.  
In Table~\ref{partialwidth}, the partial width for $Z
\to \U \gamma$ is tabulated with the same input parameters. The partial
width is only sizable for $d_\U \le 1.4$.
\begin{table}[t!]
\caption{\label{partialwidth}
Partial width of $\Gamma(Z\to \U \gamma)$ for 
$\lambda^f_1 = \lambda^{\prime f}_1=1$ and $\Lambda_\U=1$ TeV
for $d_\U = 1.1 - 2.0$.
}
\begin{tabular}{lr}
\hline
$d_\U$  &  $\;\;\;\Gamma(Z \to \U \gamma)$ (GeV) \\
\hline
\hline
1.1  &  $1.4 \times 10^{-5}$ \\
1.2  &  $8.5 \times 10^{-6}$ \\
1.3  &  $3.8 \times 10^{-6}$ \\
1.4  &  $1.5 \times 10^{-6}$ \\
1.5  &  $5.5 \times 10^{-7}$ \\
1.6  &  $1.9 \times 10^{-7}$ \\
1.7  &  $6.5 \times 10^{-8}$\\
1.8  &  $2.2 \times 10^{-8}$ \\
1.9  &  $7.0 \times 10^{-9}$ \\
2.0  &  $2.3 \times 10^{-9}$ \\
\hline
\end{tabular}
\end{table}

On the other hand, as the fermion mass inside the loop becomes infinitely heavy, one has 
the following expansion for the loop function ${\cal I}(z,\eta)$ valid for $\eta \to \infty$,
\begin{eqnarray}
{\cal I} (z, \eta) & \approx & \frac{1}{24 \eta} + \frac{ (1 + 2 z) } {360 \eta^2} \; .
\end{eqnarray}
Thus the loop amplitude $\cal A$ in Eq.(\ref{loopamp}) is vanishingly small and the heavy fermion decouples in this limit
unlike the cases in the Higgs decay of $H \to \gamma \gamma,  Z \gamma$  and $\U \gamma$.

\begin{figure}[t!]
\centering
\includegraphics[width=5.8in]{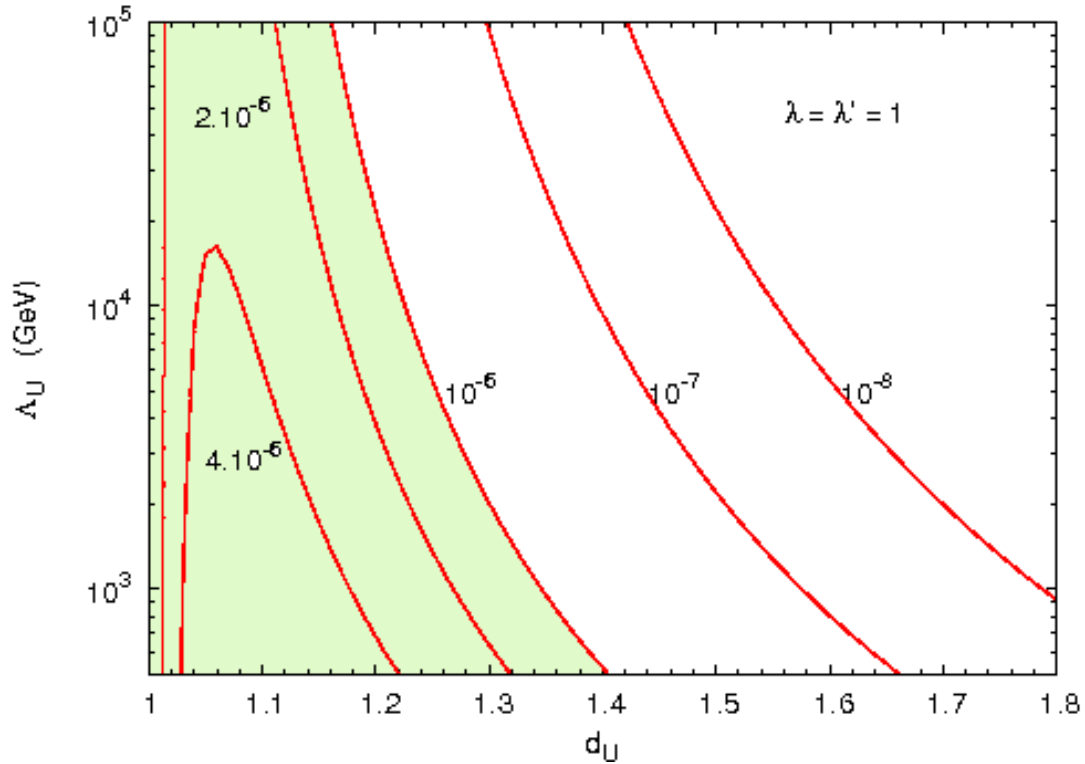}
\caption{\small\label{constraint}
Contour plot for the branching ratio $B(Z \to \U \gamma)$ versus 
($d_\U , \Lambda_\U$) with $\lambda^f_1 = \lambda^{\prime f}_1=1$.
The shaded region to the left of the contour $10^{-6}$ is ruled out by the
95\% C.L. upper limit on $B(Z \to \gamma X) < 1\times 10^{-6}$.}
\end{figure}

Experimental searchs for $e^-e^+ \to \gamma X,$ where $X$ represents a
weakly interacting stable particle, have been performed at the $Z$
resonance \cite{LEP-I-bounds}.  No signal was found, and the 95\% C.L. upper
limit on the branching ratio is \cite{LEP-I-bounds}
\begin{equation}
\label{LEP-I-bounds}
B(Z \to \gamma X) \leq 1 - 3 \times 10^{-6} \;,
\end{equation}
who's range depends on $E_{\rm min}$, the minimum energy cut  of the photon.
For $E_{\rm min}$ between 30 GeV and $m_Z/2,$ the branching
ratio upper limit is roughly a constant of about $1 \times 10^{-6}$.  It is
clear from Fig. \ref{spectrum} that in $Z\to \U \gamma ,$ most
contributions come from the region where $E_\gamma \ge 30$ GeV.
Therefore, we use the limit $B(Z \to \U \gamma ) < 1\times 10^{-6}$ to
constrain the unparticle parameter space.
In Fig.~\ref{constraint}, we plot the contour of the branching ratio
for $Z \to \U \gamma$ as a function of $d_\U$ and $\Lambda_\U$
assuming democratically as before that $\lambda^f_1 = \lambda^{\prime f}_1 = 1$
for all SM fermions.  One sees that existing limits from LEP I can already
place useful constraints on the hidden unparticle sector.

To recap, we have studied the rare decay of the $Z$ boson into a
single photon plus unparticle that has both vector and axial-vector
couplings to the SM fermions.  Existing limits from LEP I were used to
constrain the parameters of the hidden unparticle sector associated
with vector and/or axial-vector unparticle.  Despite having a peculiar
photon energy distribution in this 2-body decay, the branching ratio
is rather minuscule and at best of the order of $10^{-6}$ for small
scaling dimension $d_\U \leq 1.4 $. For larger scaling dimension, the
branching ratio is at least smaller by two orders of magnitude. Unless
one can collect a very large sample of $Z$ boson, detection of the
unparticle through this mode would be quite challenging.

\section*{Acknowledgments}
This research was supported in parts by the NSC
under grant number NSC 96-2628-M-007-002-MY3, the NCTS of Taiwan (Hsinchu) as well as 
U. S. DOE under grant numbers DE-FG02-84ER40173 and DE-FG05-85ER40226.

\end{document}